\documentclass{svproc}
\usepackage{url}
\usepackage{subcaption}
\usepackage{graphicx}
\usepackage{soul}
\usepackage[hidelinks]{hyperref}
\usepackage[square,numbers]{natbib}
\usepackage{tikz}
\usepackage{pgf}
\usepackage{subcaption}
\usepackage{pgf-umlsd}
\usetikzlibrary{arrows.meta}
\usepackage{xcolor}
\usepackage{layouts}

\usepackage[misc]{ifsym}

\begin{document}
\mainmatter              
%


\title{Analysis of Asset Administration Shell-based Negotiation Processes
for Scaling Applications}
\titlerunning{Scaling Asset Administration Shells}  
%
\author{David Dietrich \Letter \and Armin Lechler \and Alexander Verl}
\authorrunning{David Dietrich et al.} 

\institute{ISW, University of Stuttgart, Seidenstr. 36, 70174 Stuttgart, Germany\\
\email{david.dietrich@isw.uni-stuttgart.de}}

\maketitle              

\begin{abstract}
The proactive Asset Administration Shell (AAS) enables bidirectional communication between assets.
It uses the Language for I4.0 Components in VDI/VDE 2193 to facilitate negotiations, such as allocating products to available production resources.
This paper investigates the efficiency of the negotiation, based on criteria, such as message load, for applications with a scaling number of assets. 
Currently, the focus of AAS standardization is on submodels and their security to enable interoperable data access.
Their proactive behavior remains conceptual and is still a subject of scientific research.
Existing studies examine proactive AAS architecture examples with a limited number of assets, raising questions about their scalability in industrial environments.
To analyze proactive AAS for scaling applications, a scenario and evaluation criteria are introduced.
A scalable implementation is developed using current architectures for proactive AAS, upon which experiments are conducted with a varying number of assets.
The results reveal the performance limitations, communication overhead, and adaptability of the AAS-based negotiation mechanism scaling. 
This information can improve the further development and standardization of the AAS.
\end{abstract}

\section{Introduction} \label{sec:intro}

The planning of production processes typically relies on central software systems, such as manufacturing execution systems (MES).
However, decentralized systems are becoming increasingly popular due to factors such as scalability, flexibility, and data sovereignty.
Research on agent systems in the 2000s has advanced to the internet of things and digital twins with distributed intelligence and peer interaction.
The negotiation process as one possible peer interaction consists of service requester $SR$ and service provider $SP$ establishing 
a flexible horizontal cooperation for distributed tasks in a bidding procedure.
Corresponding use cases in industrial manufacturing that require decentral negotiation include plug-and-produce scenarios \cite{Bennulf.2020}  or flexible, capability-based planning \cite{Dietrich.2024c}.
Enablers for Software-defined Value Networks with data spaces such as Catena-X\footnote{https://catena-x.net/} also form the basis for interoperable negotiations along value chains and for use cases such as cross-company production planning \cite{Dietrich.2024e}.
For a successful industrial use, current technologies to realize digital twins need scalable solutions.

\subsection{Foundations} \label{subsec:found}
One standard, which attempts to realize digital twins is the Asset Administration Shell (AAS\footnote{For the sake of consistency, no distinction is made between singular and plural forms; the abbreviation \textit{``AAS''} is used uniformly throughout.})\cite{Zhang.2025}, which is specified by the Industrial Digital Twin Association \cite{IndustrialDigitalTwinAssociatione.V..2025}.
The AAS is a structured collection of an asset's data throughout its entire lifecycle and serves as a single point of access to this data.
Therefore, the specification contains a metamodel of the AAS, the definition of a REST-API, a package file format called AASX, and a security concept.
A distinction is made between the integration of the AAS, which can be passive in a file or active on a server with the possibility to communicate with its asset, or other AAS, and services.
For the communication with other AAS and services the classification of communication and presentation (CP) is used which distinguishes between a passive, active or industry 4.0 compliant communication.
The term proactive AAS is commonly used to describe an AAS with CP44 compliant industry 4.0 communication with control sovereignty over its asset or other AAS.

\begin{figure}
    \centering
    \begin{tikzpicture}[>=Latex, thick]
  \node (SR) at (0,0) {$SR_j$};
  \node (SP) at (5,0) {$SP_i$};

  \draw (0,-0.5) -- (0,-5.5);
  \draw (5,-0.5) -- (5,-5.5);

  \draw[->] (0,-1) -- (4.85,-1) node[midway, above] {call for proposal};
  \draw[->] (5,-1.5) -- (0.15,-1.5) node[midway, above] {not understood};
  \draw[->] (5,-2) -- (2.5,-2) node[midway, above] {no answer};
  \draw[->] (5,-2.5) -- (0.15,-2.5) node[midway, above] {refusal};
  \draw[->] (5,-3) -- (0.15,-3) node[midway, above] {offer};
  \draw[->] (0,-3.5) -- (4.85,-3.5) node[midway, above] {offer rejection};
  \draw[->] (0,-4) -- (4.85,-4) node[midway, above] {offer acceptance};
  \draw[->] (5,-4.5) -- (0.15,-4.5) node[midway, above] {error};
  \draw[->] (5,-5) -- (0.15,-5) node[midway, above] {conforming};

  \filldraw[fill=white] (-0.15,-0.6) rectangle (0.15,-1);
  \filldraw[fill=white] (4.85,-1) rectangle (5.15,-3);
  \filldraw[fill=white] (-0.15,-1.5) rectangle (0.15,-1.9);
  \filldraw[fill=white] (-0.15,-2.5) rectangle (0.15,-2.9);
  \filldraw[fill=white] (-0.15,-3) rectangle (0.15,-4);
  \filldraw[fill=white] (4.85,-3.5) rectangle (5.15,-3.9);
  \filldraw[fill=white] (4.85,-4) rectangle (5.15,-5);
  \filldraw[fill=white] (-0.15,-4.5) rectangle (0.15,-4.9);
  \filldraw[fill=white] (-0.15,-5) rectangle (0.15,-5.4);

\end{tikzpicture}
    \caption{Single-stage bidding procedure   \cite{VereinDeutscherIngenieure.c}}
    \label{fig:negotiation-process}
\end{figure}
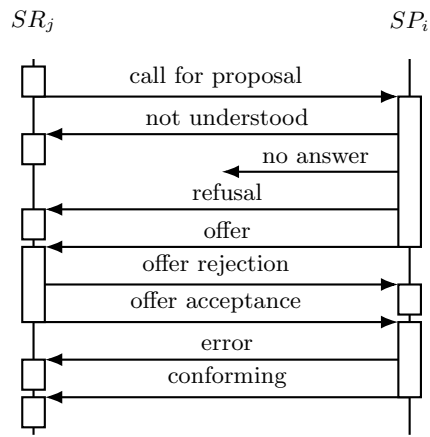

The CP44-compliant I4.0 communication between a service requester $SR$ and service provider $SP$ is specified in \cite{VereinDeutscherIngenieure., VereinDeutscherIngenieure.c} and displayed in \autoref{fig:negotiation-process}.
The specification is subdivided into vocabulary, the structure of message, and the interaction protocols, which, for instance, enable negotiation processes like bidding procedures.
However, it lacks a formal definition for use in software tools or communication protocols.

\subsection{Related Work} \label{subsec:related}
The concept of proactive AAS is increasingly mentioned alongside the specifications in current research.
Although numerous applications use the AAS as a data model to implement business logic and interaction capabilities, only few works implement these as proactive AAS according to the standards.
\cite{Belyaev.2019b} presents an architecture for an AAS that extends the passive part with a component manager, algorithms, and an interaction manager to realize the communication capabilities and logic.
A model of proactive behavior with a corresponding state machine is presented in \cite{Schroder.2019}.
In \cite{Grunau.2022b}, a division of the functionalities between the AAS server and a separate application, as well as the infrastructure required for communication, is discussed.
Different approaches to realize the architecture, for example, monolithic or agent-based, as well as interfaces for the interaction aligned with the specified REST interface are described in \cite{Stolze.2024b}.
A bidding process for a production planning scenario is introduced in \cite{Stolze.2024} using business process models.

It is remarkable that, throughout the described evaluation use cases in the literature, only prototypical implementations with a handful of AAS are considered.
Even though the theoretical approaches of the architectures concern scaling, an evaluation of these, as well as a scaling of the interaction approaches to a large number of assets and thus the suitability for an industrial scale has not yet been considered.
This arises the research question: How can current approaches of proactive AAS be scaled and what performance do they have regarding their interaction - for example, in terms of negotiation time or message load?

To answer this research question, this work introduces a modeling of the proactive AAS system in \autoref{sec:scale}.
\autoref{sec_res} presents the analyzed scenario with its implementation and results.
A discussion of these is carried out in \autoref{sec_disc} and concluded in \autoref{sec:conc}.

\section{Modeling of the Proactive AAS System} \label{sec:scale}

This section addresses metrics and methodologies for describing and analyzing a scalable system of proactive AAS.
The system consists of $m$ service providers ($SP_i$) and $n$ service requester ($SR_j$) which define roles in the negotiation process.
For the negotiation of the service a description, boundary conditions, and optimization criteria are extracted from the introduced literature.
Regarding the multiple interactions after another its execution needs to be included in the modeling.
For simplicity, each $SP$ refers to a set of offered services described by $\Sigma_i^{SP} \subseteq \Sigma$ and each $SR$ refers to a single requested service $\sigma_j^{SR} \in \Sigma$ for a set of semantically described services $\Sigma = { \sigma_1, \sigma_2, \dots, \sigma_K}$.
This approach neglects a hierarchical composition and extensive service modeling.
Other attributes of a $SP$ are the costs $c_i$, the process duration $d_i$, and the delivered quality level $q_i^{out}$.
For a $SR$, the attributes are the budget $b_j$, 
and the quality requirement $q_j^{req}$.
Regarding the interaction of messages, each $SP$ and $SR$ has an expected response time $t_{i}^{SP,exp}$ or $t_{j}^{SR,exp}$ specified in \cite{VereinDeutscherIngenieure.}.

A service provider $SP_i$ is considered capable $\kappa_{ij}^{c} = \sigma_j^{SR} \in \Sigma_i^{SP} \land (q_j^{req} \leq q_i^{out})$ of processing a service request $j$ if the descriptions and quality levels match.
In addition, it is feasible $\kappa_{ij}^{f} = \kappa_{ij}^{c} \land (c_i \leq b_j)$ when budget and quality are met.
The interaction is considered actionable $\kappa_{ij}^{a} = \kappa_{ij}^{c} \land (t_{now} - t_i^{last} > t^{exp}) \land (t_{now} - t_{iu}^{agr} > d_u)$ if no other offer is provided within the expected response time and if the work of the last agreement with preceding requester, $u$, is performed.
Due to the condition in \cite{VereinDeutscherIngenieure.}, that an offer is binding, only an actionable service provider may respond an offer to a call for proposals (CFP).
The service requester in turn collects offers within $t^{exp}$ and chooses the best according to costs $c_i$.

In the modeled system, a service provider can fulfill multiple requests after another.
To analyze the negotiation process, $SP$ and $SR$ are initialized at $t_{start}$ and simulated simultaneously.
Each $SR$ counts its CFP $\chi_j^{cfp}$, and its success with resulting costs $c_{\nu_k}$.
The relation of the $SR_k$ to the fulfilling $SP_m$ is described by the mapping $\nu_k$ with $k \in {1,2,\dots,m}$.
Furthermore, each $SR$ and $SP$ counts its messages $\chi_{SP_i}^{mes}$ and $\chi_{SR_j}^{mes}$.
The system is balanced, so the simulation ends at $t_{end}$, with a simulation time $\Delta t=t_{end}-t_{start}$, as soon as CFP of all $SR$ have been fulfilled or when no feasible providers are found.
\subsection{Criteria} \label{subsec:crit}
A generic metric to measure the scalability $\Psi$ of a distributed system is proposed by \cite{Jogalekar.1998} as:
\begin{equation} \label{eq:psi-gen}
    \Psi = F(\lambda_2,QoS_2,C_2)/(F(\lambda_1,QoS_1,C_1)).
\end{equation}
$F$ evaluates the system on two scales $1$ and $2$ regarding revenue $\lambda$, quality of service $QoS$ and costs $C$, where $F$ is proportional to $\lambda/C$.
To obtain a scaling behavior, \cite{Jogalekar.1998} introduces a scaling strategy based on the scaling factor $k$.

Applied to the AAS negotiation process, we consider revenue $\lambda=n$ as the number of service requests and costs $C=m$ as the number of service providers.

The quality of service of the system $QoS=\Delta t$ is defined as the time required to achieve a balanced system.
As the evaluation of the system shall increase on a decreasing $QoS$ the evaluation $F=\lambda/(C*QoS)$ is used.
Thus from \autoref{eq:psi-gen} results:
\begin{equation}
    \Psi = \frac{F_2}{F_1}=\frac{\lambda_2C_1QoS_1}{\lambda_1C_2QoS_2}=\frac{n_2m_1\Delta t(k_1)}{n_1m_2\Delta t(k_2)} \label{eq_psi_model}
\end{equation}

To get further insights in the system, the number of average negotiations per call for proposal $\tau^{cfp}=\sum_{j=1}^n\chi_j^{cfp}/n$ and the average messages $\tau^{mes}=\sum_{i=1}^m\chi_{SP_i}^{mes}/m+\sum_{j=1}^n\chi_{SR_j}^{mes}/n$ are also considered as metrics.
Regarding the outcome of the negotiations, we also propose the average successful requests $\tau^{s}=s/n$ for $s=(\sum_{k \in n | s_k=1}1)$ and the total costs per successful request $\tau^{c}=(\sum_{k \in n | s_k=1}c_{\nu_k})/\tau^{s}$.

\subsection{Methodology} \label{subsec:meth}
Regarding the technology-agnostic methodology of the interaction in \autoref{fig:negotiation-process}, two questions arise when systems of proactive AAS are scaled:
\begin{itemize}
    \item How can service requesters initially find out which AAS they can request?
    \item According to what principles do service providers create offers?
\end{itemize}
Regarding the first question, the three network communication methods unicast, multicast, and broadcast are considered.
It can be determined that current evaluation scenarios assume one or more known providers \cite{Stolze.2024} or publish a broadcast CFP \cite{Grunau.2022b}.
For an optimized multicast behavior, the common semantic service description $\sigma$ can be used either for a discovery service which determines capable $SP$, or as a topic in the call.

Addressing the second question, current implementations process single calls without respect to others, which resembles a first-in first-out principle regarding scalability.
Together, they enable the AAS scalability of the research question.

\section{Implementation and Results} \label{sec_res}

For the evaluation of the scalability, the following scenario, is assumed:
\begin{itemize}
    \item Factor $k$ is scaled binary logarithmic from $2^0$ to $2^8$ with twice as many service requesters $n=2k$ as service providers $m=k$, resembling different manufacturing systems. 
    By that \autoref{eq_psi_model} simplifies to $\Psi = \Delta t(k_1)/\Delta t(k_2)$.
    \item Out of different production capabilities $\Sigma=A,B,C,D,E$, each $SR_j$ is described by a random service request $\sigma_j^{SR}$, while $SP_i$ receives uniformly distributed one to three capabilities.
    \item Costs, work duration, and budget $c_i, d_i, b_i = \mathcal{N}(\mu_{c,d,b},\sigma_{c,d,b})$ are normally distributed with $\mu_c=10$, $\mu_b=1,1*\mu_c=11$ , $\sigma_c=\sigma_b=1$, $\mu_d=20s$, and $\sigma_d=2s$.
    \item The required quality level $q_j^{req}\in\{0,1,2,3\}$ reaches uniformly distributed from zero to three with provided quality level likewise from two to five.
    \item The response times $t_i^{SP,exp}=t_j^{SR,exp}=t^{exp}=5s$ are set as equal.
\end{itemize}
Hence a multicast request based on the semantic description $\sigma$ should not affect $\psi$ but only reduce the message load $\tau^{mes}$, only a multicast scenario with first-in first-out offer creation is analyzed regading scalability.
For each factor $k$ ten simulation runs will be conducted as sample size.
The generated parameters of $SP$ and $SR$ in each simulation run are modeled in a passive properties of an AAS with a unique ID.

Regarding the implementation of the negotiation process in a proactive AAS, no common standard or open-source implementation was found, but are realized individually by tools like Node-red\footnote{\url{https://nodered.org/}}, Flowable\footnote{https://www.flowable.com/} or custom programs.
As $d_i$ and $t^{exp}$ are expected to play a major role compared to runtime performance, the negotiation process is implemented in Python with MQTT for communication between the assets.
The message structure is compliant with \cite{VereinDeutscherIngenieure.} and the respective interactive elements needed from the above model as topics ``cfp/$\sigma_j$'' for a multicast CFP and ``message/AAS-ID'' for responses to an existing conversation based on the ID of the target AAS.
Each $SR$ and $SP$ with its logic, the interaction manager, an MQTT-client as messenger, and the passive AAS compliant to \cite{IndustrialDigitalTwinAssociatione.V..2025}, are encapsulated in a respective docker container building a proactive AAS.
An orchestration service conducts the scaling strategy using replicas in a docker swarm and collects the results.

\begin{figure}[htbp]
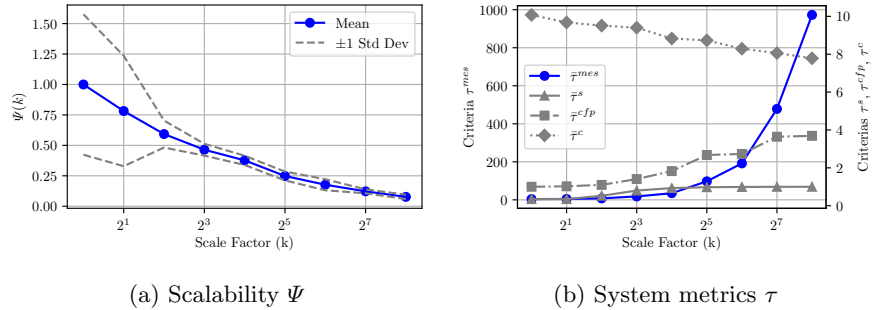

    \centering
    \begin{subfigure}[b]{0.48\linewidth}
        \centering
        \resizebox{\textwidth}{!}{\input{figures/Results_phi.pgf}}
        \caption{Scalability $\Psi$}
        \label{subfig:psi}
    \end{subfigure}
    \begin{subfigure}[b]{0.48\linewidth}
        \centering
        \resizebox{\textwidth}{!}{\input{figures/Results_taus.pgf}}
        \caption{System metrics $\tau$}
        \label{subfig:tau}
    \end{subfigure}
    \caption{Results the negotiation process for scaling strategy with factors $k$}
    \label{fig:results}
\end{figure}

The scenario was conducted on a virtual machine in Proxmox with Ubuntu 24.04, 90GB RAM, 32 x86-64-v4 processors on a single docker swarm manager node.
Its results regarding the criteria in \autoref{subsec:crit} are displayed in \autoref{fig:results}.
\autoref{subfig:psi} shows the mean $\Psi$ and its standard deviation over the ten simulation runs per $k$ on a binary logarithmic scale with $\bar{\Delta t}(k_1)=21.7s$.
According to \cite{Jogalekar.1998}, it resembles a scaling behavior which is unscalable.
Regarding the other criteria in \autoref{subfig:tau}, an exponential behavior in the average message load $\bar{\tau}^{mes}$ can be observed.
The average success rate $\bar{\tau}^s$ starts at almost zero and approaches nearly one at a scale of $2^3$ $SP$.
The average CFP $\bar{\tau}^{cfp}$ increases from about one between $k=1$ and $k=2^3$ to four at $k=2^7$.
After a peak over $9.5$ at $k=1$ the average costs $\bar{\tau}^{c}$ decrease to eight.

\section{Discussion} \label{sec_disc}

This section gives a brief discussion of the previously presented results with respect to threats to their validity and possible improvements.
$\bar{\tau}^s$ shows a low success rate for scale factors lower than $2^3$ and results from the random distribution of capabilities and their quality leading to a partial mismatch between $SR$ and $SP$.
The average CFP supports this thesis, as with around one CFP only few requests are scheduled after a first one where a $SP$ was capable but not actionable.
The increasing average CFP also shows, that for higher $k$ a $SR$ needs more attempts to find actionable providers.
This may be due to the fact, that a $SR$ chooses the cost-best proposal leading to a queue for other $SR$ with less budget.
This cost reduction can also be observed at $\bar{\tau}^c$.
The increasing message load $\bar{\tau}^{mes}$ resembles the increasing communication overhead of even the multicast communication with plenty of answers for each CFP and may lead to network limitations for even a leightweight MQTT communication at scale.
These metrics are also reflected in the unscalable behavior of $\Psi$ with simulation times up to $360s$.
In comparison, the duration of in average two requests per provider including duration and response time would lead to about $30s$.

A major influence of the performance may be the orchestration, especially when at the maximum $n_{k=8}=2\cdot 2^8=512$ service providers are started simultaneously leading to a delayed behavior.
Nevertheless, all $512$ $SR$ are started within a timeout of $90s$, making a significant influence on $\Psi$ but keeping its trend clearly intact.
Further performance limitations and thus threats to validity may lie in the custom implementation like the cost optimal selection or message processing, but are expected as minor role due to the observed response time.
Compared to industrial use, simplification in the service description are made.
The methods to scale systems of proactive AAS from \autoref{subsec:meth} based on multicast CFP and first-in first-out offer provision have been demonstrated successfully.

The results show various fields of action to further improve scalability of proactive AAS.
On the one hand the negotiation behavior of the proactive AAS needs further respect.
This includes the questions raised in \autoref{subsec:meth} representing the common standard in the analysis.
Besides the cost optimal selection of $SR$, also the provision of binding offers of $SP$ needs further attention.
For example, ingoing CFP within the response time may also be cached and the offer is provided for a best fit.
On the other hand the scalability can be improved by architecture patterns known of decentralized software systems, such as supernodes, zones or service meshes.
They could enable a subdivision of $SR$ and $SP$ at scale and support message flow and decision logic by mapping cached CFP to best-fit offers across zones at a system level.
\section{Conclusion} \label{sec:conc}

This work presented an analysis of the negotiation process, as part of the industry 4.0 language in VDI/VDE 2193, at scale.
It shows, that performance limitations are to be expected, when current methodologies tested with few proactive AAS are scaled to industrial manner with hundreds of assets.
With regard to the research question of how proactive AAS approaches can be scaled, a scalable behavior of $SR$ and $SP$ for CFP and offers was designed.
Furthermore, a modeling for the proactive AAS system and target criteria to measure scalability are developed.
The implementation based on a scalable docker swarm with corresponding results are introduced and discussed.

However, several aspects are identified to future work improving scalability.
First, regarding the negotiation process, the behavior of proactive AAS and architectural patterns of the system need to be improved for a better negotiation an scheduling.
Second, the implementation of the business logic and algorithms in a proactive AAS, which in this case was done in a custom python implementation, needs to be addressed in the further standardization.
Last, the interaction elements of I4.0 Messages and their semantics need further attention for interoperable bidding procedures.

\newblock

\noindent \textbf{Acknowledgment.} This research and development project is funded by the German Federal Ministry of Research, Technology and Space (BMFTR) within the ``Research Campus – Public-Private Partnership for Innovation'' funding initiative (02P23Q820 ff) and managed by the Project Management Agency Karlsruhe (PTKA). The authors are responsible for the content of this publication.
\newline
\renewcommand{\bibsection}{%
  \noindent\textbf{\Large Literature}
}

\bibliographystyle{spbasic_unsrt}
\bibliography{bibliography}

\end{document}